\documentclass{aa}
\usepackage{graphicx}
\begin{document}

\title{Precise Reddening and Metallicity of NGC6752 from FLAMES spectra
\thanks{Based on data collected at the European Southern Observatory with
the VLT-UT2, Paranal, Chile (ESO 073.D-0100)}}

\author{R.G.~Gratton\inst{1},
        A.~Bragaglia\inst{2},
        E.~Carretta\inst{2},
        F. De Angeli\inst{3},
        S.~Lucatello\inst{1},
        Y.~Momany\inst{3},
        G.~Piotto\inst{3},
        A.~Recio Blanco\inst{3,4}}

\offprints{R.G.~Gratton}

\institute{INAF-Osservatorio Astronomico di Padova, Vicolo
        dell'Osservatorio 5, 35122 Padova, ITALY
\and
        INAF-Osservatorio Astronomico di Bologna, Via Ranzani 1,
        40127 Bologna, ITALY
\and
        Dipartimento di Astronoma, Universit\'a di Padova, Vicolo
        dell'Osservatorio 5,35122 Padova, ITALY
\and    
        Observatoire Astronomique de la C\^ote d'Azur,
        Boulevard de l'Observatoire, B.P. 4229, F-06304 Nice Cedex 4 	}

\date{Received: 08 March 2005; accepted: 04 May 2005}

\abstract{
Accurate reddenings for Globular Clusters could be obtained by comparing
the colour-temperature obtained using temperatures from reddening-free
indicator ($H\alpha$), with that given by standard colour-temperature
calibrations. The main difficulty in such derivations is the large errors in
temperatures for individual stars due to uncertainties on the removal of
instrumental signature for each individual star. The large multiplexing
opportunity offered by FLAMES at VLT2 allowed us to obtain spectra centred on
$H\alpha$\ at a resolution of R=6000 and $5<S/N<50$\ for 120 stars near the
turn-off of NGC6752 with GIRAFFE from a single 1300 seconds exposure. This set
of spectra was used to derive effective temperatures from fittings of
$H\alpha$\ profiles with typical errors of about $\pm 200$~K and reddening
estimates with individual errors of 0.05 mag. Averaging all individual
reddenings, a high precision reddening estimate has been obtained for the
cluster: $E(B-V)=0.046\pm 0.005$. The same exposure provided UVES spectra of
seven stars near the red giant branch bump at a resolution of 40,000, and
$20<S/N<40$. These spectra, combined with temperatures from colours
(corrected for our high precision reddening value) provided Fe abundances with
internal errors of 0.026~dex, and with average metallicity [Fe/H]=$-1.48\pm
0.01\pm 0.06$~dex (random + systematic). Abundances were obtained for several
other elements, allowing e.g. an accurate estimate of the ratio between the
$\alpha-$elements and Fe ([$\alpha$/Fe]=$+0.27\pm 0.01$). The O-Na
anticorrelation is evident from our UVES data, in agreement with past
results.\\
This analysis shows the power of FLAMES for analysis of globular clusters: the
accurate reddenings and metal abundances obtained by a procedure like that
described here, combined with distance determinations from cluster dynamics or
main sequence fitting, and high quality colour-magnitude diagrams, could allow
derivation of ages with errors below 1 Gyr for individual globular clusters.
}

\authorrunning{Gratton R.G. et al.}
\titlerunning{Reddening and Metallicity of NGC6752}

\maketitle

%

\section{INTRODUCTION}

Accurate estimates of the absolute ages of (the oldest) Globular Clusters
(GCs), coupled with determinations of the Hubble constants $H_0$\ from the
spectrum of fluctuations of the microwave background determined from the WMAP
experiment (Spergel et al. 2003), or from the HST Key Project (Freedman et al.
2001), may provide a stringent lower limit to the age of the Universe and
constrains the exponent $w$\ of the equation of state of the dark energy
(Jimenez et al. 2003), independently of type Ia SN observations (see e.g.
Gratton et al. 2003a).

Furthermore, within the framework of a standard $\Lambda$CDM model, where the
age of the Universe is accurately fixed at $13.7\pm 0.2$~Gyr by the WMAP
results (Spergel et al. 2003), the age of GCs can be used to constrain the
epoch of formation of the Galaxy, linking the local Universe to the distant
one (see Carretta et al. 2000; and Gratton et al. 2003a).

Relative ages are fundamental to describe the early history of our Galaxy. In
this framework it should be noted that galactic GCs divide into two main
groups: halo and thick disk (or bulge) GCs (Zinn 1985). The differential ages
method suggests that these two groups might have ages different by about 2 Gyr
(Rosenberg et al. 1999); this result seems supported by absolute ages (Gratton
et al. 2003a). This has implications for both cosmology (where only the oldest
GCs are of interest) and galactic evolution. Observations of GC systems in
other galaxies suggest a link between GC formation and strong dynamical
interactions (Peebles \& Dicke 1968; Schweizer \& Seitzer 1993). The oldest
group of GCs might then be related to the very early phases of the galactic
collapse, while the second one may instead trace a later accretion event (see
Freeman \& Bland-Hawthorn, 2002), possibly related to the end of the thick
disk phase indicated by chemistry (Gratton et al. 1996, Gratton et al. 2000;
Fuhrmann 1998).

An important goal is then to derive absolute ages with internal errors of $\pm
1$~Gyr for an ample sample of GCs. Ages for GCs with such small errors may be
derived only using the luminosity of the turn-off (TO): this on turn requires
accurate distances, with errors $<5\%$. In perspective, most accurate and
robust distances (error $<2$\%) for a few GCs will be obtained using
geometrical methods (Piotto et al 2004). At present, distances with errors of
3-5\% can be obtained for a larger sample of GCs using the Main Sequence
Fitting Method (MSFM), exploiting local subdwarfs as standard candles (see
Gratton et al. 1997; Pont et al. 1998; Carretta et al. 2000; Gratton et al.
2003a). Main sources of errors in MSFM are possible systematic differences in
reddenings and metallicities between field and GC stars (accurate initial He
abundances in GCs have been determined by Cassisi et al. 2003; see also
Salaris et al. 2004). Both of them can be reduced to within the required
accuracy if a reddening-free temperature indicator is used for both field
and GC stars of similar evolutionary phases: the analysis of the results of
the ESO LP 165.L-0263 (Gratton et al. 2003a) relative to three GCs (NGC6397,
NGC6752 and 47~Tuc) spanning almost the total metallicity range of galactic
GCs showed that this approach may provide reddenings accurate to
$\Delta(B-V)=\pm 0.005$~mag, metallicities accurate to $\pm 0.04$~dex,
distances accurate to 4\%, and ages with errors of about $\pm 1$~Gyr. Also
geometrical distances (which determines the true distance modulus toward a
cluster) will take advantage from accurate reddening and metallicity
determinations, since apparent distance moduli are required to derive ages.

Note that here we are only interested in relative reddening and metallicity
determinations: the adopted scale may be tied to that of field stars
exploiting the clusters observed within the LP 165.L-0263 (47 Tuc, NGC6397,
and NGC6752: Gratton et al. 2003a). The same temperature indicator may be
adopted (H$\alpha$\ profile).

In this paper we describe a pilot program on NGC6752 which exploits the
multiplexing capabilities of FLAMES, the VLT multifibre facility (Pasquini et 
al. 2002). The large number of spectra that could be obtained using GIRAFFE 
allowed a proper reduction of the major source of errors in temperatures 
derived from H$\alpha$: flat fielding. On the other side, low resolution and 
S/N were not too critical in such observations, allowing to use faint turn-off 
stars. The simultaneous acquisition of spectra of a few red giants with UVES 
allowed additionally an accurate determination of the chemical composition. 
For this purpose we preferred to use relatively warm stars, for which the 
analysis is expected to be robust. The success of this procedure suggests the 
usefulness of an extensive program on other globular clusters. Ages within
$\pm 1$~Gyr are now fully within reach for a substantial sample of them in
both Zinn's groups.

\begin{table}
\begin{center}
\caption{Summary of observations}
\begin{tabular}{lc}
\hline
Date                 & 24/06/2004 \\
Time (UT Start)      & 09:49:51   \\
Exposure Time (sec)  &    1300    \\
Airmass (Mean)       &   1.803    \\
Seeing FWHM (arcsec) &    0.62    \\ 
\hline
\end{tabular}
\label{t:summary_obs}
\end{center}
\end{table}

\section{OBSERVATIONS}

Data used in this paper are based on a single 1300 seconds exposure obtained
on June 24th, 2004, with FLAMES at Kueyen (=VLT2) used in service mode. The
observations were obtained at a rather large airmass (about 1.8) and in very
good seeing conditions (FWHM=0$\farcs$62 at zenith and 5000~\AA) (see also
Table~\ref{t:summary_obs}).

120 fibres feeding the GIRAFFE spectrograph were centred on stars slightly
brighter than the turn-off of NGC6752, in the magnitude range $16.7<V<17.2$.
Stars were carefully selected from high quality photometric $UBV$ 
observations obtained with the Wide-Field Imager (WFI) at the 2.2~m ESO-MPI 
telescope (total field  of view of $34\times33$ arcmin$^2$). For a detailed 
representation of the data reduction and calibration of this data set we refer 
the reader to Momany et al. (2004).

The astrometric calibration of  NGC6752 reference images employed over 7000
stars from the GSC2.2 catalogue (Loomis et al. 2004), using the IRAF MSCRED
package\footnote{IRAF is distributed by the National Optical Astronomy
Observatory, which is operated by the Association of Universities for Research
in Astronomy, Inc, under cooperative agreement with the National Science
Foundation.}. The internal accuracy of the astrometry has been estimated to be
about 0.15 arcsec, well within the  FLAMES requirements (0.2 arcsec). To
further confirm the fulfilment of the requirements, we matched our
astrometrically calibrated NGC6752 catalogue with that of UCAC2 (Zacharias et
al. 2004) and estimated the positional residuals for 550 stars in common. The
residuals show a Gaussian distribution with an r.m.s. of $\simeq$0.05 arcsec
in both coordinates.

Only uncrowded stars were considered, that is stars not showing any companion
brighter than $V_{\rm target}+2$~mag within 2.5~arcsec, or brighter than
$V_{\rm target}-2$~mag within 10~arcsec. The targets were selected to lie
close to the cluster mean loci in the colour-magnitude diagram. A posteriori,
radial velocities confirmed membership of all but two of the observed stars.
Eight fibres were additionally used to monitor sky background; they were
pointed toward carefully selected empty sky regions.

The GIRAFFE spectrograph was used with the LR06 grating; the spectra cover the
wavelength range 6400-7100~\AA\ at a resolution of about $R\sim 6000$.
Pixel-to-pixel S/N of the spectra (measured from the scatter of individual
spectral points in the wavelength range 6660-6670~\AA) ranged from 5 to 50,
with typical values around 20. The S/N values were generally lower for stars
in the outer regions of the cluster, most likely because these stars were not
well centred on the fibre heads. This can be attributed to the effects of
differential refraction at the rather large airmass of observation.

The average of the eight sky spectra were used to subtract telluric emissions
(in particular, emission in H$\alpha$) from GIRAFFE spectra. Appropriate
scaling factors were evaluated to take into account the transmission of
individual fibres.

Seven fibres feeding the UVES spectrograph were centred on stars close
to the RGB bump ($13.2<V<14.2$), while one was dedicated to the sky. 
The spectra cover the wavelength range 4700-6900~\AA, and have $20<S/N<40$. 

The two sets of spectra were reduced using the dedicated FLAMES pipelines 
(BLDRS Python software 0.5.3 version for the GIRAFFE spectra; uves/2.1.1
version for  UVES spectra). We found that this UVES pipeline does not
accurately subtract the background between orders in the green-yellow part of
the spectra. Only a few lines measured on these portions of the spectra were
considered in the present analysis.

\section{REDDENING ESTIMATES FROM GIRAFFE SPECTRA}

\subsection{Fluxes}

The following procedure was used to derive accurate temperatures from the
H$\alpha$\ profile. First, instrumental fluxes within 10 narrow bands of
5~\AA\ width in the region including $H\alpha$\ were measured on the GIRAFFE
spectra by integrating the observed spectra, after shifting them in wavelength
for the geocentric radial velocity of each star. Cosmic ray hits were removed
before evaluating the fluxes. The list of the bands used is given in
Table~\ref{t:bands}.

\begin{table}
\begin{center}
\caption{Definition of the bands used to derive T$_{\rm eff}$'s}
\begin{tabular}{lcc}
\hline
Band & Start & End   \\
     & (\AA) & (\AA) \\
\hline
1 & 6537.8 & 6542.8 \\ 
2 & 6542.8 & 6547.8 \\
3 & 6547.8 & 6552.8 \\
4 & 6552.8 & 6557.8 \\
5 & 6557.8 & 6562.8 \\
6 & 6562.8 & 6567.8 \\
7 & 6567.8 & 6572.8 \\
8 & 6572.8 & 6577.8 \\
9 & 6577.8 & 6582.8 \\
10& 6582.8 & 6587.8 \\
\hline
\end{tabular}
\label{t:bands}
\end{center}
\end{table}

The fluxes measured in each band were then normalized to a {\it
pseudocontinuum} given by a straight line connecting the average of the first
two bands with the average of the two last bands. The normalized fluxes for
all the stars observed with GIRAFFE are given in Table~\ref{t:fluxes}
(available only in electronic form). The second column of this Table gives
also the $S/N$\ ratio for each spectrum, computed from the spectral region
6660-6670~\AA, where there is no significant feature.

Since the $H\alpha$\ profiles are expected to be fairly symmetric and since
bands are defined symmetrically with respect to the line center, errors in
these normalized fluxes can be obtained by comparing fluxes measured on the
blue and red side of the $H\alpha$. Eliminating a few outliers, the
comparisons are as follows:
\begin{equation}
5-6 = -0.008\pm 0.002,~~~~{\rm r.m.s.}=0.019,~~~{\rm 112~stars}
\end{equation}
and:
\begin{equation}
4-7 = -0.013\pm 0.002,~~~~{\rm r.m.s.}=0.020,~~~{\rm 114~stars}
\end{equation}
for bands 5-6 and 4-7. From these comparison, we expect a typical error in
the average of $<5,6>$\ bands of $\pm 0.0095$, corresponding to an internal
error in T$_{\rm eff}$'s of $\pm 110$~K (see next Section). 

This error agrees with expectations based on the $S/N$\ of the
spectra.

Individual heliocentric radial velocities measured by the FLAMES pipeline are
presented in the third column of Table ~\ref{t:fluxes}.  In a few cases (7
stars out of 120), these measures of the radial velocities by the automatic
routine in the pipeline were obviously wrong, perhaps due to the strong
telluric signal present in this wavelength range. Radial velocities for these
stars were measured by  fitting H$\alpha$, and zeroing the radial velocity on
the telluric bands consistently with the other stars. All but one (29049) of
the stars appear to be members of the cluster on the basis of radial velocity;
the velocity for star 39462 is discrepant too, but this spectrum has very low
S/N and we suspend judgement about it. However we excluded these two stars
from our estimates of reddening of NGC6752. The mean radial velocity is
$-32.0\pm 0.6$~km~s$^{-1}$, with an r.m.s. scatter of individual values of
6.0~km~s$^{-1}$ (118 stars), with no obvious correlation of the spread with
S/N, nor distance from the cluster centre.

The average value of the radial velocity agrees very well with those estimated
by  Webbink (1988: $-32.2\pm 3.2$~km~s$^{-1}$) and Dubath et al. (1997:
$-32.0\pm 1.6$~km~s$^{-1}$), while it is slightly larger than that given by
Rutledge et al. (1997: $-27.4\pm 2.7$~km~s$^{-1}$). For comparison, the seven
stars observed with UVES provided a slightly lower average velocity ($-23.8\pm
2.1$~km~s$^{-1}$, see Table~\ref{t:uvesphot}).

Errors in the GIRAFFE radial velocities should be roughly 5-6 km
s$^{-1}$, as given by the FLAMES pipeline. For comparison, the radial velocity
error is expected to be roughly $\sigma(RV)\sim 33/(S/N)$~km s$^{-1}$ when
using the formula by Landman et al. (1982) and assuming that all radial
velocity signal is given by H$\alpha$ alone: for the typical $S/N$\ of the
program spectra, errors are then expected in the range 0.7-6~km s$^{-1}$.
Given these large uncertainties on the errors attached to these radial
velocities, they are of little use in estimating the internal velocity
dispersion in NGC~6752.

\begin{table*}
\caption{S/N ratios, heliocentric radial velocities and normalized fluxes for
the target stars (Table to appear only in electronic form)}
\begin{tabular}{lccccccccc}
\hline
Star  &$S/N$&$V_r$&   3   &   4   &   5   &   6   &   7   &   8   &$<5,6>$\\
      &     &km s$^{-1}$ &       &       &       &       &       &       &       \\
\hline
 1538 & 18 & $-$34.0 & 0.008 & 0.042 & 0.273 & 0.291 & 0.073 & 0.008 & 0.282 \\
 1895 & 35 & $-$35.6 & 0.021 & 0.043 & 0.217 & 0.242 & 0.039 & 0.029 & 0.229 \\
 2119 & 24 & $-$35.4 & 0.015 & 0.047 & 0.218 & 0.253 & 0.053 & 0.024 & 0.236 \\
 2250 & 30 & $-$28.8 & 0.024 & 0.069 & 0.243 & 0.251 & 0.061 & 0.012 & 0.247 \\
 2374 & 36 & $-$42.2 & 0.021 & 0.038 & 0.293 & 0.167 & 0.033 & 0.010 & 0.230 \\
 2484 & 28 & $-$33.1 & 0.020 & 0.031 & 0.216 & 0.240 & 0.032 & 0.009 & 0.228 \\
 2486 & 29 & $-$31.4 & 0.005 & 0.023 & 0.207 & 0.224 & 0.035 & 0.008 & 0.215 \\
 2533 & 35 & $-$28.3 & 0.010 & 0.030 & 0.222 & 0.244 & 0.030 & 0.019 & 0.233 \\
 2541 & 18 & $-$27.9 & 0.027 & 0.070 & 0.265 & 0.286 & 0.076 & 0.013 & 0.276 \\
 2594 & 30 & $-$30.6 & 0.003 & 0.035 & 0.217 & 0.231 & 0.044 & 0.013 & 0.224 \\
 2661 & 31 & $-$31.4 & 0.008 & 0.030 & 0.210 & 0.199 & 0.036 & 0.018 & 0.204 \\
 2700 & 25 & $-$42.8 & 0.009 & 0.037 & 0.285 & 0.163 & 0.022 &-0.001 & 0.224 \\
 2750 & 27 & $-$29.3 & 0.014 & 0.040 & 0.251 & 0.241 & 0.028 & 0.019 & 0.246 \\
 4606 & 11 & $-$39.7 & 0.025 & 0.017 & 0.212 & 0.239 & 0.087 & 0.033 & 0.225 \\
 4693 & 27 & $-$34.9 & 0.036 & 0.045 & 0.278 & 0.185 & 0.036 & 0.026 & 0.231 \\
 4905 & 27 & $-$21.7 & 0.015 & 0.034 & 0.234 & 0.197 & 0.042 & 0.017 & 0.215 \\
 5540 & 18 & $-$21.0 & 0.023 & 0.042 & 0.250 & 0.263 & 0.078 & 0.038 & 0.257 \\
 6057 & 26 & $-$23.1 & 0.023 & 0.032 & 0.226 & 0.242 & 0.033 & 0.016 & 0.234 \\
 6439 & 27 & $-$27.9 & 0.026 & 0.036 & 0.221 & 0.242 & 0.039 & 0.013 & 0.232 \\
 8265 & 16 & $-$38.9 & 0.018 & 0.062 & 0.291 & 0.310 & 0.091 & 0.037 & 0.301 \\
12600 & 29 & $-$28.7 & 0.016 & 0.048 & 0.249 & 0.247 & 0.059 & 0.022 & 0.248 \\
20454 & 33 & $-$30.7 & 0.023 & 0.050 & 0.277 & 0.165 & 0.050 & 0.026 & 0.221 \\
21163 & 33 & $-$37.5 & 0.010 & 0.031 & 0.211 & 0.237 & 0.047 & 0.016 & 0.224 \\
23507 & 23 & $-$25.6 & 0.015 & 0.049 & 0.230 & 0.254 & 0.042 & 0.023 & 0.242 \\
23697 & 27 & $-$38.1 & 0.021 & 0.054 & 0.247 & 0.247 & 0.054 & 0.024 & 0.247 \\
24300 & 33 & $-$34.9 & 0.029 & 0.048 & 0.223 & 0.250 & 0.054 & 0.007 & 0.236 \\
25443 & 31 & $-$23.2 & 0.031 & 0.041 & 0.233 & 0.240 & 0.069 & 0.043 & 0.237 \\
25518 & 54 & $-$18.8 & 0.015 & 0.036 & 0.266 & 0.227 & 0.031 & 0.005 & 0.247 \\
26038 & 29 & $-$31.2 & 0.024 & 0.054 & 0.305 & 0.168 & 0.056 & 0.020 & 0.236 \\
26672 & 25 & $-$43.8 & 0.007 & 0.031 & 0.188 & 0.290 & 0.039 & 0.004 & 0.239 \\
27210 & 30 & $-$37.9 & 0.012 & 0.081 & 0.239 & 0.248 & 0.057 & 0.019 & 0.244 \\
27611 & 26 & $-$22.3 & 0.019 & 0.035 & 0.223 & 0.228 & 0.062 & 0.017 & 0.226 \\
28023 & 34 & $-$38.0 & 0.017 & 0.064 & 0.264 & 0.274 & 0.078 & 0.028 & 0.269 \\
28047 & 40 & $-$35.3 & 0.015 & 0.025 & 0.210 & 0.226 & 0.027 & 0.016 & 0.218 \\
28440 & 28 & $-$38.2 & 0.001 & 0.041 & 0.219 & 0.240 & 0.040 & 0.018 & 0.230 \\
28598 & 36 & $-$33.5 & 0.015 & 0.048 & 0.254 & 0.280 & 0.053 &-0.014 & 0.267 \\
29049 & 39 &  ~~20.8 & 0.024 & 0.056 & 0.248 & 0.230 & 0.063 & 0.026 & 0.239 \\
29206 & 35 & $-$33.5 & 0.007 & 0.035 & 0.226 & 0.244 & 0.051 & 0.018 & 0.235 \\
29380 & 27 & $-$38.8 & 0.006 & 0.036 & 0.222 & 0.246 & 0.042 & 0.017 & 0.234 \\
29767 & 22 & $-$29.8 & 0.020 & 0.050 & 0.231 & 0.235 & 0.060 & 0.019 & 0.233 \\
29778 & 29 & $-$40.0 & 0.016 & 0.032 & 0.210 & 0.226 & 0.042 & 0.027 & 0.218 \\
29827 & 32 & $-$28.1 & 0.004 & 0.033 & 0.210 & 0.251 & 0.026 & 0.032 & 0.230 \\
29986 & 20 & $-$31.6 & 0.017 & 0.079 & 0.276 & 0.292 & 0.075 & 0.016 & 0.284 \\
30188 & 18 & $-$34.5 & 0.015 & 0.059 & 0.259 & 0.274 & 0.081 &-0.003 & 0.266 \\
30275 & 30 & $-$20.9 & 0.019 & 0.067 & 0.279 & 0.218 & 0.066 & 0.020 & 0.249 \\
30584 & 22 & $-$33.0 & 0.036 & 0.049 & 0.232 & 0.236 & 0.053 & 0.030 & 0.234 \\
30616 & 35 & $-$33.2 & 0.004 & 0.023 & 0.215 & 0.221 & 0.048 & 0.036 & 0.218 \\
30716 & 31 & $-$31.6 & 0.028 & 0.057 & 0.242 & 0.262 & 0.061 & 0.025 & 0.252 \\
30733 & 44 & $-$23.2 & 0.032 & 0.042 & 0.263 & 0.241 & 0.075 & 0.048 & 0.252 \\
30852 & 25 & $-$44.9 & 0.005 & 0.057 & 0.250 & 0.276 & 0.080 & 0.054 & 0.263 \\
30991 & 31 & $-$29.4 & 0.033 & 0.053 & 0.220 & 0.251 & 0.050 & 0.014 & 0.235 \\
31194 & 21 & $-$26.5 & 0.041 & 0.041 & 0.259 & 0.263 & 0.058 & 0.046 & 0.261 \\
31370 & 25 & $-$33.6 & 0.014 & 0.034 & 0.266 & 0.281 & 0.051 & 0.015 & 0.273 \\
31407 & 34 & $-$29.0 & 0.025 & 0.029 & 0.210 & 0.211 & 0.039 & 0.031 & 0.210 \\
31412 & 19 & $-$27.6 & 0.032 & 0.042 & 0.255 & 0.241 & 0.075 & 0.048 & 0.248 \\
31430 & 21 & $-$30.1 & 0.035 & 0.062 & 0.249 & 0.240 & 0.072 & 0.019 & 0.245 \\
31589 & 13 & $-$28.5 & 0.040 & 0.008 & 0.271 & 0.280 & 0.085 & 0.006 & 0.276 \\
31620 & 16 & $-$38.1 & 0.036 & 0.058 & 0.243 & 0.267 & 0.072 & 0.023 & 0.255 \\
31711 & 31 & $-$33.4 & 0.008 & 0.037 & 0.231 & 0.232 & 0.044 & 0.017 & 0.231 \\
31757 & 17 & $-$36.6 & 0.043 & 0.059 & 0.284 & 0.285 & 0.061 & 0.038 & 0.284 \\
\hline
\end{tabular}
\label{t:fluxes}
\end{table*}

\addtocounter{table}{-1}

\begin{table*}
\caption{Normalized Fluxes (cont. in electronic form)}
\begin{tabular}{lccccccccc}
\hline
Star  &$S/N$&$V_r$&   3   &   4   &   5   &   6   &   7   &   8   &$<5,6>$\\
      &     &km s$^{-1}$ &       &       &       &       &       &       &       \\
\hline
31771 & 30 & $-$36.2 & 0.008 & 0.042 & 0.213 & 0.220 & 0.062 & 0.022 & 0.216 \\
31773 & 21 & $-$20.2 & 0.022 & 0.040 & 0.237 & 0.238 & 0.045 & 0.001 & 0.237 \\
31921 & 13 & $-$29.8 & 0.015 &-0.006 & 0.236 & 0.214 &-0.025 &-0.040 & 0.225 \\
31945 & 21 & $-$38.1 & 0.056 & 0.030 & 0.212 & 0.240 & 0.065 & 0.028 & 0.226 \\
32027 & 21 & $-$39.9 & 0.020 & 0.028 & 0.213 & 0.230 & 0.090 & 0.014 & 0.222 \\
32621 & 18 & $-$30.4 & 0.017 & 0.023 & 0.221 & 0.220 & 0.050 & 0.022 & 0.221 \\
32883 & 10 & $-$32.7 &-0.007 & 0.020 & 0.239 & 0.217 & 0.039 &-0.005 & 0.228 \\
33450 & 22 & $-$35.6 & 0.019 & 0.017 & 0.233 & 0.255 & 0.012 &-0.000 & 0.244 \\
33673 & 18 & $-$26.5 & 0.024 & 0.052 & 0.212 & 0.247 & 0.064 & 0.010 & 0.229 \\
33766 &  6 & $-$36.6 & 0.024 & 0.106 & 0.258 & 0.265 & 0.029 & 0.083 & 0.262 \\
33816 & 26 & $-$22.5 &-0.004 & 0.019 & 0.208 & 0.220 & 0.033 & 0.010 & 0.214 \\
34048 & 20 & $-$19.5 & 0.017 & 0.028 & 0.229 & 0.190 & 0.048 & 0.019 & 0.209 \\
34076 & 13 & $-$28.7 & 0.005 & 0.038 & 0.241 & 0.231 & 0.023 & 0.012 & 0.236 \\
34141 & 19 & $-$29.0 &-0.012 & 0.008 & 0.208 & 0.201 & 0.012 &-0.011 & 0.204 \\
34329 &  9 & $-$32.7 & 0.018 &-0.035 & 0.243 & 0.236 & 0.059 & 0.063 & 0.240 \\
34451 & 11 & $-$33.2 & 0.010 & 0.026 & 0.228 & 0.251 & 0.057 & 0.040 & 0.239 \\
34596 & 21 & $-$30.9 & 0.001 & 0.041 & 0.230 & 0.279 & 0.054 & 0.019 & 0.254 \\
34628 & 16 & $-$30.1 & 0.043 & 0.028 & 0.216 & 0.240 & 0.060 & 0.031 & 0.228 \\
34676 & 18 & $-$32.7 & 0.016 & 0.020 & 0.211 & 0.238 & 0.058 &-0.002 & 0.225 \\
35072 & 15 & $-$42.0 & 0.039 & 0.031 & 0.264 & 0.280 & 0.083 & 0.035 & 0.272 \\
35084 & 11 & $-$33.9 &-0.010 & 0.051 & 0.227 & 0.239 & 0.036 &-0.040 & 0.233 \\
35152 & 12 & $-$23.7 &-0.033 & 0.003 & 0.274 & 0.278 & 0.046 & 0.028 & 0.276 \\
35324 & 13 & $-$28.8 & 0.037 &-0.003 & 0.185 & 0.180 & 0.032 &-0.024 & 0.182 \\
35571 & 24 & $-$24.8 & 0.006 &-0.004 & 0.228 & 0.222 & 0.048 &-0.005 & 0.225 \\
35798 & 20 & $-$24.9 & 0.015 & 0.033 & 0.200 & 0.198 & 0.020 & 0.003 & 0.199 \\
36063 & 22 & $-$26.5 & 0.015 & 0.051 & 0.260 & 0.260 & 0.086 & 0.026 & 0.260 \\
36087 & 18 & $-$27.1 & 0.025 & 0.045 & 0.238 & 0.230 & 0.053 & 0.035 & 0.234 \\
36384 &  7 & $-$27.0 & 0.072 & 0.115 & 0.329 & 0.298 & 0.035 & 0.067 & 0.313 \\
36451 &  7 & $-$37.3 & 0.038 &-0.022 & 0.244 & 0.222 & 0.028 & 0.003 & 0.233 \\
36732 & 18 & $-$40.8 & 0.021 & 0.083 & 0.273 & 0.282 & 0.076 & 0.016 & 0.277 \\
37054 &  9 & $-$42.8 & 0.011 &-0.059 & 0.285 & 0.240 & 0.027 & 0.047 & 0.262 \\
37448 & 30 & $-$30.6 & 0.014 & 0.023 & 0.207 & 0.236 & 0.057 & 0.010 & 0.222 \\
37493 & 20 & $-$28.1 & 0.009 & 0.044 & 0.234 & 0.256 & 0.048 & 0.016 & 0.245 \\
37571 & 13 & $-$29.9 & 0.028 & 0.058 & 0.248 & 0.291 & 0.058 & 0.004 & 0.270 \\
37655 & 31 & $-$35.7 & 0.015 & 0.042 & 0.204 & 0.227 & 0.046 & 0.020 & 0.215 \\
37831 & 25 & $-$37.5 & 0.022 & 0.051 & 0.230 & 0.241 & 0.034 & 0.032 & 0.235 \\
38619 & 19 & $-$32.3 &-0.000 & 0.041 & 0.216 & 0.222 & 0.047 & 0.027 & 0.219 \\
38652 & 20 & $-$39.8 &-0.003 & 0.023 & 0.193 & 0.238 & 0.017 &-0.008 & 0.216 \\
39026 &  6 & $-$33.6 & 0.032 & 0.075 & 0.307 & 0.315 & 0.053 & 0.061 & 0.311 \\
39064 & 16 & $-$33.9 & 0.005 & 0.041 & 0.226 & 0.262 & 0.047 & 0.007 & 0.244 \\
39255 & 14 & $-$28.7 & 0.048 & 0.030 & 0.273 & 0.302 & 0.098 & 0.050 & 0.287 \\
39379 & 19 & $-$37.3 & 0.029 & 0.040 & 0.240 & 0.278 & 0.084 & 0.054 & 0.259 \\
39391 & 18 & $-$31.1 & 0.019 & 0.047 & 0.247 & 0.245 & 0.058 & 0.015 & 0.246 \\
39451 & 15 & $-$39.8 & 0.018 & 0.042 & 0.258 & 0.326 & 0.068 & 0.010 & 0.292 \\
39462 &  5 & $-$61.5 & 0.040 &-0.002 & 0.318 & 0.294 &-0.015 & 0.055 & 0.306 \\
39554 & 11 & $-$32.1 & 0.024 & 0.031 & 0.255 & 0.255 & 0.062 & 0.027 & 0.255 \\
39559 & 10 & $-$24.0 &-0.014 & 0.122 & 0.295 & 0.267 & 0.112 & 0.068 & 0.281 \\
39612 &  5 & $-$39.2 & 0.100 & 0.097 & 0.299 & 0.318 & 0.153 & 0.028 & 0.308 \\
39775 & 21 & $-$40.1 & 0.022 & 0.036 & 0.219 & 0.237 & 0.050 & 0.013 & 0.228 \\
39984 & 12 & $-$21.2 & 0.022 & 0.049 & 0.239 & 0.248 & 0.114 & 0.052 & 0.244 \\
40147 &  6 & $-$29.6 & 0.045 & 0.125 & 0.289 & 0.289 & 0.051 & 0.002 & 0.289 \\
46655 & 12 & $-$29.5 & 0.056 & 0.033 & 0.245 & 0.280 & 0.051 & 0.009 & 0.263 \\
48532 & 11 & $-$19.8 & 0.017 & 0.027 & 0.256 & 0.233 & 0.038 & 0.007 & 0.245 \\
48884 &  7 & $-$40.0 & 0.061 & 0.043 & 0.267 & 0.289 & 0.043 & 0.010 & 0.278 \\
48905 & 14 & $-$38.9 & 0.008 & 0.072 & 0.235 & 0.226 & 0.075 & 0.033 & 0.231 \\
48916 & 11 & $-$30.7 & 0.000 & 0.051 & 0.259 & 0.247 & 0.079 & 0.035 & 0.253 \\
49005 & 12 & $-$38.6 & 0.017 & 0.029 & 0.233 & 0.223 & 0.038 & 0.036 & 0.228 \\
49143 & 11 & $-$31.7 &-0.004 & 0.001 & 0.245 & 0.258 & 0.017 & 0.003 & 0.252 \\
49311 & 17 & $-$38.5 &-0.005 & 0.013 & 0.247 & 0.255 & 0.064 & 0.037 & 0.251 \\
49342 & 17 & $-$36.4 & 0.022 & 0.006 & 0.222 & 0.226 & 0.052 & 0.026 & 0.224 \\
\hline
\end{tabular}
\label{t:fluxes}
\end{table*}

\begin{figure}
\includegraphics[width=8.8cm]{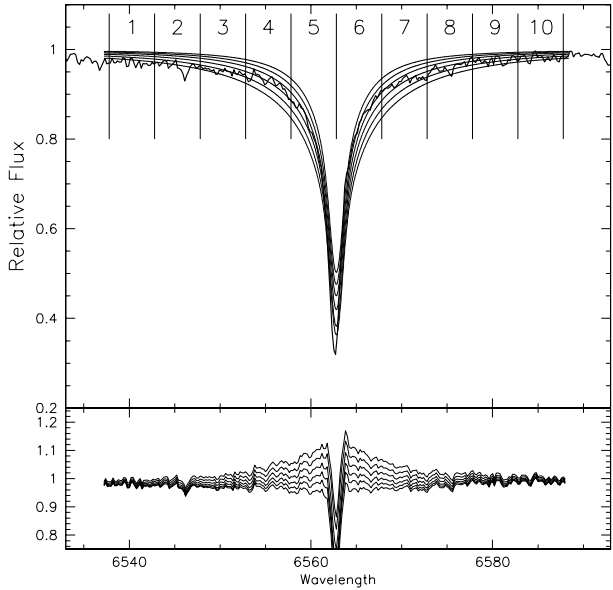}
\caption[]{Top panel: Average H$\alpha$\ spectrum for the stars with Giraffe spectra with $S/N>25$. Overimposed are synthetic spectra computed for gravities and metal abundances appropriate for the program stars, and various values of $T_{\rm eff}$\ (=5600, 5800, 6000, 6200, 6400, 6600~K). The limits of the bands used are also shown as vertical marks. Bottom panel: ratios between observed and synthetic spectra.}
\label{f:halfa}
\end{figure}

\subsection{Effective temperatures}

Effective temperatures were derived by comparing the normalized average fluxes
in bands 5 and 6 with those expected from Kurucz (1992) model atmospheres
(with the overshooting option switched off) of different temperatures.
Gravities and metal abundances adopted for these model atmospheres are those
appropriate for the program stars. The $H\alpha$\ absorption and broadening
were modelled using the same assumptions of Castelli et al. (1997). The
theoretical profiles were further broadened by convolution with a Gaussian
profile mimicking the instrumental profile.

To put these effective temperatures derivation in a more clear perspective, we
show in Fig.~\ref{f:halfa} the average spectrum for the stars with Giraffe
spectra with $S/N>25$\ in the region around H$\alpha$. Overimposed are
synthetic spectra computed for gravities and metal abundances appropriate for
the program stars, and various values of T$_{\rm eff}$\ (=5600, 5800, 6000,
6200, 6400, 6600~K). The limits of the bands used are also shown as vertical
marks.

Values for the T$_{\rm eff}$'s are given in Table ~\ref{t:teff} (available
only in electronic form).

\begin{figure}
\includegraphics[width=8.8cm]{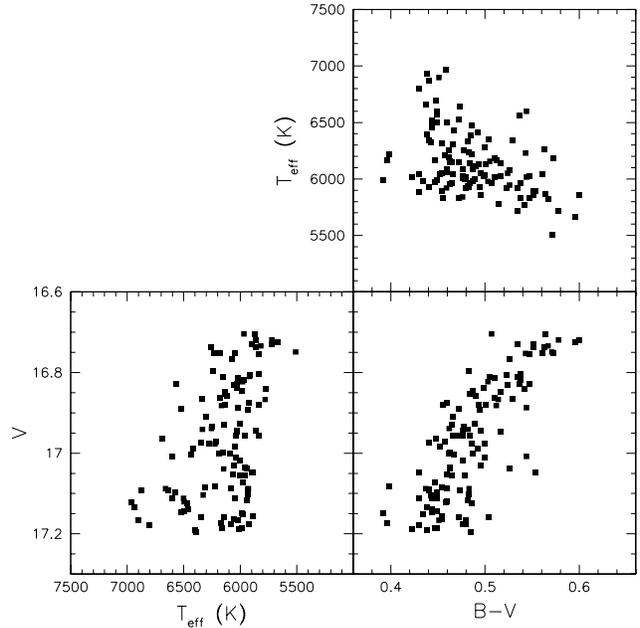}
\caption[]{Top panel:  Effective Temperature - $B-V$\ colour diagram for the
stars observed with GIRAFFE; Left panel: Effective Temperature - $V$\
magnitude diagram for the same stars; right panel: $B-V$\ colour - $V$\
magnitude diagram for the same stars.} \label{f:vteff}
\end{figure}

\section{Photometry and reddening}

Reddening estimates for individual stars can be derived by comparing the
observed colours with those predicted from the reddening-free temperatures and
an appropriate colour-temperature relation.

The $B-V$\ colours were given by the WFI photometry (Momany et al. 2004). They
were corrected blueward by 0.020 mag to put them on the same scale of Thompson
et al. (1999) used in Gratton et al. (2003a). Based on the scatter around the
mean relation, we expect that errors in colours for individual stars are of
$\pm 0.025$~mag at this magnitude.

We found clear correlations between the T$_{\rm eff}$'s derived from
$H\alpha$, the $B-V$\ colours and the $V$\ magnitudes within our sample (see
Figure~\ref{f:vteff}).

\begin{table}
\begin{center}
\caption{Temperatures and Reddenings (Table to appear only in electronic form)}
\begin{tabular}{lcccc}
\hline
Star  &   V    &  B-V  &T$_{\rm eff}$ & E(B-V) \\ 
      &        &       & (K)          &        \\
\hline
 1538 & 17.097 & 0.449 &6575 & ~~0.092 \\
 1895 & 17.073 & 0.447 &5975 &$-$0.026 \\
 2119 & 16.983 & 0.452 &6045 &$-$0.004 \\
 2250 & 17.174 & 0.396 &6168 &$-$0.033 \\
 2374 & 16.847 & 0.487 &5986 & ~~0.016 \\
 2484 & 17.056 & 0.465 &5960 &$-$0.012 \\
 2486 & 16.881 & 0.455 &5832 &$-$0.056 \\
 2533 & 16.887 & 0.544 &6016 & ~~0.081 \\
 2541 & 17.143 & 0.449 &6498 & ~~0.081 \\
 2594 & 16.810 & 0.535 &5917 & ~~0.047 \\
 2661 & 16.731 & 0.535 &5720 &$-$0.008 \\
 2700 & 16.806 & 0.523 &5918 & ~~0.035 \\
 2750 & 16.998 & 0.463 &6155 & ~~0.032 \\
 4606 & 17.101 & 0.483 &5935 &$-$0.001 \\  
 4693 & 17.055 & 0.478 &6000 & ~~0.011 \\
 4905 & 16.805 & 0.538 &5833 & ~~0.027 \\
 5540 & 16.977 & 0.500 &6277 & ~~0.094 \\
 6057 & 16.944 & 0.476 &6028 & ~~0.016 \\
 6439 & 17.017 & 0.476 &6005 & ~~0.010 \\
 8265 & 17.179 & 0.430 &6801 & ~~0.098 \\
12600 & 16.864 & 0.510 &6182 & ~~0.085 \\
20454 & 17.156 & 0.454 &5889 &$-$0.042 \\
21163 & 17.175 & 0.479 &5923 &$-$0.008 \\
23507 & 16.858 & 0.490 &6115 & ~~0.050 \\
23697 & 17.185 & 0.447 &6166 & ~~0.018 \\
24300 & 17.031 & 0.495 &6053 & ~~0.041 \\
25443 & 17.164 & 0.454 &6055 & ~~0.000 \\
25518 & 16.882 & 0.512 &6167 & ~~0.083 \\
26038 & 16.832 & 0.524 &6052 & ~~0.069 \\
26672 & 16.768 & 0.526 &6077 & ~~0.077 \\
27210 & 16.879 & 0.493 &6131 & ~~0.056 \\
27611 & 17.117 & 0.482 &5938 &$-$0.001 \\
28023 & 16.988 & 0.492 &6414 & ~~0.110 \\
28047 & 16.721 & 0.600 &5859 & ~~0.096 \\
28440 & 17.150 & 0.434 &5981 &$-$0.038 \\
28598 & 17.191 & 0.438 &6396 & ~~0.053 \\
29049 & 17.175 & 0.476 &6084 & ~~0.029 \\
29206 & 17.012 & 0.500 &6035 & ~~0.041 \\
29380 & 16.839 & 0.499 &6029 & ~~0.039 \\
29767 & 16.813 & 0.510 &6019 & ~~0.047 \\
29778 & 16.738 & 0.551 &5859 & ~~0.047 \\
29827 & 17.186 & 0.449 &5989 &$-$0.021 \\
29986 & 17.111 & 0.449 &6599 & ~~0.095 \\
30188 & 17.196 & 0.485 &6389 & ~~0.099 \\
30275 & 17.002 & 0.461 &6188 & ~~0.037 \\
30584 & 16.947 & 0.517 &6026 & ~~0.056 \\
30616 & 16.943 & 0.495 &5861 &$-$0.008 \\
30716 & 16.977 & 0.486 &6223 & ~~0.069 \\
30733 & 16.753 & 0.543 &6230 & ~~0.127 \\  
30852 & 17.158 & 0.504 &6349 & ~~0.111 \\
30991 & 16.956 & 0.466 &6040 & ~~0.009 \\
31194 & 17.104 & 0.443 &6328 & ~~0.046 \\
31370 & 17.125 & 0.486 &6469 & ~~0.113 \\
31407 & 16.867 & 0.514 &5780 &$-$0.012 \\
31412 & 16.751 & 0.572 &6180 & ~~0.146 \\
31430 & 17.158 & 0.472 &6144 & ~~0.038 \\
31589 & 17.120 & 0.460 &6497 & ~~0.091 \\
31620 & 16.738 & 0.563 &6259 & ~~0.153 \\
31711 & 16.926 & 0.489 &6000 & ~~0.022 \\
31757 & 17.008 & 0.544 &6599 & ~~0.190 \\
\hline
\end{tabular}
\label{t:teff}
\end{center}
\end{table}

\addtocounter{table}{-1}

\begin{table}
\begin{center}
\caption{Temperature and Reddenings (cont.)  (Table to appear only in electronic form)}
\begin{tabular}{lcccc}
\hline
Star  &   V    &  B-V  &T$_{\rm eff}$ & E(B-V) \\ 
      &        &       & (K)          &        \\
\hline
31771 & 16.956 & 0.476 &5841 &$-$0.033 \\
31773 & 17.053 & 0.460 &6064 & ~~0.008 \\
31921 & 17.088 & 0.483 &5929 &$-$0.002 \\
31945 & 17.039 & 0.526 &5946 & ~~0.045 \\
32027 & 17.049 & 0.554 &5897 & ~~0.061 \\
32621 & 17.048 & 0.430 &5887 &$-$0.066 \\
32883 & 16.819 & 0.538 &5967 & ~~0.063 \\
33450 & 16.849 & 0.516 &6137 & ~~0.081 \\
33673 & 16.823 & 0.503 &5979 & ~~0.031 \\
33766 & 16.941 & 0.482 &6334 & ~~0.086 \\
33816 & 16.734 & 0.567 &5818 & ~~0.052 \\
34048 & 16.840 & 0.542 &5771 & ~~0.013 \\
34076 & 16.753 & 0.561 &6047 & ~~0.105 \\
34141 & 16.719 & 0.578 &5721 & ~~0.035 \\
34329 & 17.002 & 0.487 &6089 & ~~0.041 \\
34451 & 17.087 & 0.459 &6084 & ~~0.012 \\
34596 & 16.934 & 0.477 &6252 & ~~0.066 \\
34628 & 16.957 & 0.484 &5963 & ~~0.007 \\
34676 & 17.089 & 0.441 &5928 &$-$0.044 \\
35072 & 17.138 & 0.444 &6458 & ~~0.069 \\
35084 & 17.166 & 0.479 &6018 & ~~0.016 \\
35152 & 17.112 & 0.444 &6502 & ~~0.076 \\
35324 & 16.749 & 0.571 &5509 &$-$0.040 \\
35571 & 16.893 & 0.494 &5929 & ~~0.009 \\
35798 & 16.724 & 0.596 &5668 & ~~0.037 \\
36063 & 17.084 & 0.454 &6314 & ~~0.055 \\
36087 & 16.828 & 0.547 &6028 & ~~0.087 \\
36384 & 17.121 & 0.458 &6964 & ~~0.139 \\
36451 & 17.187 & 0.422 &6015 &$-$0.041 \\
36732 & 17.146 & 0.444 &6518 & ~~0.079 \\
37054 & 16.866 & 0.529 &6339 & ~~0.134 \\
37448 & 16.731 & 0.552 &5897 & ~~0.059 \\
37493 & 17.039 & 0.463 &6147 & ~~0.030 \\
37571 & 17.003 & 0.467 &6431 & ~~0.088 \\
37655 & 16.956 & 0.472 &5832 &$-$0.039 \\
37831 & 17.111 & 0.430 &6043 &$-$0.027 \\
38619 & 16.705 & 0.564 &5871 & ~~0.063 \\
38652 & 16.755 & 0.547 &5834 & ~~0.036 \\
39026 & 17.134 & 0.438 &6935 & ~~0.117 \\
39064 & 16.853 & 0.484 &6137 & ~~0.049 \\
39255 & 17.093 & 0.473 &6638 & ~~0.124 \\
39379 & 16.910 & 0.466 &6304 & ~~0.065 \\
39391 & 16.812 & 0.505 &6156 & ~~0.074 \\
39451 & 16.963 & 0.448 &6691 & ~~0.105 \\
39462 & 17.092 & 0.441 &6873 & ~~0.115 \\
39554 & 16.939 & 0.462 &6255 & ~~0.051 \\
39559 & 16.828 & 0.537 &6563 & ~~0.178 \\
39612 & 17.166 & 0.451 &6898 & ~~0.127 \\
39775 & 17.035 & 0.463 &5958 &$-$0.015 \\
39984 & 16.881 & 0.501 &6132 & ~~0.065 \\
40147 & 17.088 & 0.437 &6660 & ~~0.090 \\
46655 & 16.974 & 0.440 &6345 & ~~0.046 \\
48532 & 16.930 & 0.465 &6144 & ~~0.031 \\
48884 & 16.889 & 0.472 &6525 & ~~0.108 \\
48905 & 17.149 & 0.392 &5990 &$-$0.078 \\
48916 & 16.797 & 0.483 &6239 & ~~0.069 \\
49005 & 16.704 & 0.507 &5967 & ~~0.032 \\
49143 & 17.082 & 0.398 &6221 &$-$0.019 \\
49311 & 16.971 & 0.457 &6213 & ~~0.038 \\
49342 & 16.876 & 0.459 &5920 &$-$0.028 \\
\hline
\end{tabular}
\label{t:teff}
\end{center}
\end{table}

Reddenings toward NGC6752 can be finally evaluated by comparing the observed
colour-temperature relation with that expected from models. Each individual
star provided a reddening estimate. They are listed in the last column of
Table~\ref{t:teff}. No clear trend in these reddening estimates with e.g. $V$\
magnitude or location on the field could be discerned . The individual values
were then averaged together to provide a best estimate. Values obtained in
this way are listed in Table~\ref{t:reddening}. For comparison, the value
obtained by using a similar procedure from spectra of 20 stars taken at higher
resolution (Gratton et al.2003a) is also given, as well as the values from the
compilation by Harris (1996), and from the reddening maps of Schlegel et al.
(1998).

There are several aspects in this procedure that may introduce systematic
errors in our reddening derivations (systematic uncertainties in flat
fielding, inappropriate modelling of $H\alpha$, errors in the model
atmospheres, photometric errors etc.). However, we remind here that what
matters for the age derivations are not the absolute values of the reddenings,
but rather the relative values between cluster and field stars. The good
agreement with the determination of Gratton et al. (2003a), that was on a
scale consistent with that adopted for the field subdwarfs, supports then the
present technique.

\begin{table}
\begin{center}
\caption{Various Reddening Estimates}
\begin{tabular}{lcc}
\hline
Estimates                     &    E(B-V)        &   rms   \\
                              &     mag          &   mag   \\
\hline
All stars (118)               & $0.046\pm 0.005$ &  0.053  \\
Only stars with $S/N>15$ (90) & $0.042\pm 0.005$ &  0.049  \\
                              &                  &         \\ 
Gratton et al. (2003a)        & $0.040\pm 0.005$ &         \\
Harris (1996)                 &     0.04         &         \\
Schlegel et al. (1998)        &     0.056        &         \\
\hline
\end{tabular}
\label{t:reddening}
\end{center}
\end{table}

\section{METALLICITY FROM UVES SPECTRA}

\subsection{Equivalent Widths}

\begin{table}
\begin{center}
\caption{Photometry and spectrum data for stars observed with UVES}
\begin{tabular}{lccccc}
\hline
Star  &   V    & B-V   & RV & S/N & $\sigma$(EW) \\
      &        &       & km s$^{-1}$    &    & (m\AA) \\
\hline
25072 & 14.199 & 0.803 & $-$26.4 & 28  &  3.8 \\
26059 & 13.444 & 0.853 & $-$17.7 & 40  &  2.6 \\
30409 & 13.568 & 0.840 & $-$22.9 & 44  &  2.7 \\
30426 & 14.075 & 0.802 & $-$30.0 & 28  &  3.1 \\
34854 & 13.698 & 0.814 & $-$15.5 & 19  &  5.8 \\
37999 & 13.273 & 0.866 & $-$25.3 & 32  &  2.9 \\
39672 & 13.662 & 0.825 & $-$29.1 & 23  & 10.6 \\
\hline
\end{tabular}
\label{t:uvesphot}
\end{center}
\end{table}

Table~\ref{t:uvesphot} gives the main parameters for the stars observed with
UVES. The colours have been corrected blueward by 0.008 mag to put them onto
the same scale of Thompson et al. (1999). Note that this correction is
slightly different from that found appropriate for the TO-stars; this suggests
the presence of a colour term in one of the two photometry.

The EWs were measured on the spectra using the ROSA code (Gratton 1988; see
Table~\ref{t:ewidths}) with Gaussian fittings to the measured profiles: these
exploit a linear relation between EWs and FWHM of the lines, derived from a
subset of lines characterized by cleaner profiles. Since the observed stars
span a very limited parameter range, errors in these EWs can be computed by
comparing values derived from individual stars with the average value for the
whole sample. Typical errors obtained using this procedure are listed in the
last column of Table~\ref{t:uvesphot}. They are roughly reproduced by the
formula $\sigma$(EW)$\sim 100/(S/N)$~m\AA. Considering the resolution and
sampling of the spectra, the errors in the EWs are in agreement with
expectations based on photon noise statistics (Cayrel 1988). Finally, we
notice that due to the problems in background subtraction in the green-yellow
part of the spectra, only lines with wavelength $>5900$~\AA\ were considered,
save for Na, Mg and Si, for which also lines in the 5600-5750~\AA\ region were
considered.

\begin{table*}
\caption{Equivalent Widths from UVES spectra (in electronic form)}
\begin{tabular}{lccccccccc}
\hline
Wavel.  &E.P.  &log gf &25072 &26529 &30409 &30426 &34854 &37999 &39672 \\
(\AA)   &(eV)  &       &(m\AA)&(m\AA)&(m\AA)&(m\AA)&(m\AA)&(m\AA)&(m\AA)\\
\hline
        &      &         &      &      &      &      &      &      &      \\
Fe I    &      &         &      &      &      &      &      &      &      \\
        &      &         &      &      &      &      &      &      &      \\
5930.19 & 4.65 & $-$0.29 & 41.4 & 29.7 & 33.0 & 34.3 & 26.5 & 39.0 & 51.1 \\
5934.67 & 3.93 & $-$1.15 & 29.6 & 29.1 &      & 28.4 &      & 33.8 &      \\
5956.71 & 0.86 & $-$4.60 & 33.7 & 38.5 & 35.6 & 30.2 & 40.7 & 44.6 &      \\
5984.83 & 4.73 & $-$0.39 & 27.4 &      &      &      &      & 27.3 &      \\
6003.02 & 3.88 & $-$1.08 & 35.0 & 33.0 & 31.5 & 34.2 & 31.6 & 47.5 & 56.6 \\
6008.57 & 3.88 & $-$0.96 & 32.3 & 38.9 & 31.5 & 31.8 & 38.4 & 43.9 &      \\
6027.06 & 4.08 & $-$1.23 &      &      &      &      &      & 28.2 &      \\
6065.49 & 2.61 & $-$1.53 & 87.4 & 88.6 & 89.1 & 72.0 & 90.4 & 99.6 &104.6 \\
6137.00 & 2.20 & $-$2.95 & 35.6 & 45.6 & 44.3 & 34.5 &      & 47.1 & 46.9 \\
6151.62 & 2.18 & $-$3.30 &      &      &      &      &      & 40.0 &      \\
6173.34 & 2.22 & $-$2.88 & 42.5 & 53.7 & 44.4 & 40.0 & 43.0 & 56.8 & 45.1 \\
6200.32 & 2.61 & $-$2.44 & 38.7 & 48.2 & 42.5 & 37.0 & 39.2 & 50.2 &      \\
6213.44 & 2.22 & $-$2.54 & 66.5 & 71.0 & 60.9 & 56.8 & 60.7 & 74.7 & 73.9 \\
6219.29 & 2.20 & $-$2.43 & 71.2 & 71.7 & 66.6 & 66.2 & 64.6 & 78.1 & 67.8 \\
6232.65 & 3.65 & $-$1.22 & 36.4 & 44.5 & 44.4 & 35.1 & 28.4 & 49.9 &      \\
6240.65 & 2.22 & $-$3.23 &      & 27.5 &      &      &      & 32.9 &      \\
6246.33 & 3.60 & $-$0.73 & 62.9 & 70.4 & 60.1 & 63.1 & 55.2 & 72.1 & 51.5 \\
6252.56 & 2.40 & $-$1.69 & 88.5 & 96.2 & 94.6 & 87.5 & 98.8 &103.5 &106.5 \\
6265.14 & 2.18 & $-$2.55 & 67.9 & 75.8 & 68.1 & 64.6 & 77.6 & 80.9 & 72.7 \\
6270.23 & 2.86 & $-$2.46 &      &      &      &      &      & 32.3 &      \\
6297.80 & 2.22 & $-$2.74 & 49.9 & 57.5 & 52.2 & 51.2 & 40.6 & 62.6 & 49.6 \\
6301.51 & 3.65 & $-$0.72 & 62.0 & 66.6 & 63.6 & 54.5 & 58.9 & 69.1 & 52.4 \\
6322.69 & 2.59 & $-$2.43 & 42.7 & 52.3 & 50.8 & 43.3 & 46.1 & 53.1 &      \\
6335.34 & 2.20 & $-$2.27 & 76.9 & 82.5 & 77.5 & 74.5 & 74.6 & 86.8 & 83.5 \\
6411.66 & 3.65 & $-$0.60 & 67.4 & 77.8 & 74.1 & 72.0 & 78.6 & 81.8 & 66.3 \\
6421.36 & 2.28 & $-$2.03 & 81.3 & 89.3 & 86.8 & 85.0 & 88.3 & 92.2 & 85.5 \\
6481.88 & 2.28 & $-$2.98 & 29.0 & 42.0 & 38.8 & 33.8 & 33.5 & 45.6 &      \\
6498.94 & 0.96 & $-$4.70 &      & 33.4 & 33.1 & 28.4 &      & 40.5 &      \\
6593.88 & 2.43 & $-$2.42 & 49.5 & 62.3 & 62.6 & 58.8 & 50.0 & 62.9 & 54.7 \\
6609.12 & 2.56 & $-$2.69 & 32.5 & 33.7 & 36.8 &      & 30.0 & 38.5 &      \\
6750.16 & 2.42 & $-$2.62 & 35.7 & 67.9 & 49.2 & 42.7 & 54.1 & 49.5 &      \\
        &      &         &      &      &      &      &      &      &      \\
Fe II   &      &         &      &      &      &      &      &      &      \\
        &      &         &      &      &      &      &      &      &      \\
6247.56 & 3.89 & $-$2.33 &      & 33.4 & 27.4 & 28.9 &      & 32.6 &      \\
6432.68 & 2.89 & $-$3.58 &      & 30.9 &      &      &      & 32.0 &      \\
6456.39 & 3.90 & $-$2.10 & 37.2 & 42.0 & 43.0 & 41.1 & 41.1 & 46.2 &      \\
        &      &         &      &      &      &      &      &      &      \\
O I     &      &         &      &      &      &      &      &      &      \\
        &      &         &      &      &      &      &      &      &      \\
6300.31 & 0.00 & $-$9.75 &  3.3 &  5.9 & 14.1 &  6.4 &  9.3 & 24.2 &$<$5.0\\
        &      &         &      &      &      &      &      &      &      \\
Na I    &      &         &      &      &      &      &      &      &      \\
        &      &         &      &      &      &      &      &      &      \\
5682.65 & 2.10 & $-$0.67 & 42.2 & 56.0 & 25.4 & 29.7 & 31.9 &      &      \\
5688.22 & 2.10 & $-$0.37 & 56.6 & 78.1 & 56.0 & 60.7 & 59.2 & 42.8 & 74.0 \\
        &      &         &      &      &      &      &      &      &      \\
Mg I    &      &         &      &      &      &      &      &      &      \\
        &      &         &      &      &      &      &      &      &      \\
5528.42 & 4.34 & $-$0.52 &130.7 &145.0 &139.1 &142.4 &155.1 &146.3 &169.1 \\ 
5711.09 & 4.34 & $-$1.73 & 53.3 & 46.0 & 68.2 & 42.6 & 65.4 & 64.4 & 48.3 \\
        &      &         &      &      &      &      &      &      &      \\
Si I    &      &         &      &      &      &      &      &      &      \\
        &      &         &      &      &      &      &      &      &      \\
5684.49 & 4.95 & $-$1.65 & 27.7 &      & 35.3 & 24.6 & 33.1 & 28.4 &      \\ 
5690.43 & 4.95 & $-$1.87 &      &      &                           & 27.7 \\
5708.40 & 4.95 & $-$1.47 & 37.4 & 29.1 & 40.2 & 27.8 & 42.7 & 34.8 &      \\
5948.55 & 5.08 & $-$1.23 & 34.3 & 32.6 & 30.2 & 30.3 & 36.9 & 33.0 & 43.0 \\
        &      &         &      &      &      &      &      &      &      \\
\hline
\end{tabular}
\label{t:ewidths}
\end{table*}

\addtocounter{table}{-1}

\begin{table*}
\caption{Equivalent Widths from UVES spectra (cont. in electronic form)}
\begin{tabular}{lccccccccc}
\hline
Wavel.  &E.P.  &log gf &25072 &26529 &30409 &30426 &34854 &37999 &39672 \\
(\AA)   &(eV)  &       &(m\AA)&(m\AA)&(m\AA)&(m\AA)&(m\AA)&(m\AA)&(m\AA)\\
\hline
        &      &         &      &      &      &      &      &      &      \\
Ca I    &      &         &      &      &      &      &      &      &      \\
        &      &         &      &      &      &      &      &      &      \\
6161.30 & 2.52 & $-$1.27 &      &      &      &      &      & 27.7 &      \\
6163.75 & 2.52 & $-$1.29 &      & 26.7 & 30.9 & 29.2 &      & 38.6 &      \\
6166.44 & 2.52 & $-$1.14 & 28.7 & 40.2 &      & 29.9 &      & 40.1 &      \\
6169.04 & 2.52 & $-$0.80 & 51.0 & 52.4 & 50.1 & 47.3 & 29.4 & 52.0 & 48.0 \\
6169.56 & 2.52 & $-$0.48 & 70.0 & 71.2 & 69.6 & 62.6 & 65.4 & 80.4 & 78.2 \\
6439.08 & 2.52 &  ~~0.39 &113.3 &120.0 &121.1 &122.5 &122.5 &123.5 &113.6 \\
6449.82 & 2.52 & $-$0.50 & 64.4 & 69.1 & 68.4 & 71.7 & 69.1 & 68.8 & 71.1 \\
6455.60 & 2.52 & $-$1.29 &      &      &      &      & 29.6 &      &      \\
6471.67 & 2.52 & $-$0.69 & 51.1 & 63.9 & 62.8 & 56.5 & 62.5 & 65.8 &      \\
6493.79 & 2.52 & $-$0.11 & 83.2 & 93.2 & 92.7 & 92.9 &107.4 & 98.3 & 95.0 \\
6499.65 & 2.52 & $-$0.82 & 45.6 & 55.4 & 54.6 & 45.9 & 55.8 & 54.8 &      \\
6572.80 & 0.00 & $-$4.32 &      &      & 31.5 &      &      & 31.1 &      \\
6717.69 & 2.71 & $-$0.52 & 50.8 & 70.3 & 67.9 & 55.1 & 57.9 & 66.3 & 50.7 \\
        &      &         &      &      &      &      &      &      &      \\
Sc II   &      &         &      &      &      &      &      &      &      \\
        &      &         &      &      &      &      &      &      &      \\
6245.62 & 1.51 & $-$1.05 &      & 34.9 & 27.6 &      &      & 31.4 &      \\
6279.74 & 1.50 & $-$1.16 & 28.9 &      &      & 45.7 &      & 27.5 & 51.4 \\
6604.60 & 1.36 & $-$1.15 &      & 29.6 &      &      & 34.9 &      &      \\
        &      &         &      &      &      &      &      &      &      \\
Ti I    &      &         &      &      &      &      &      &      &      \\
        &      &         &      &      &      &      &      &      &      \\
6258.11 & 1.44 & $-$0.36 & 34.8 & 36.1 & 29.7 & 29.9 &      & 43.2 &      \\
6261.11 & 1.43 & $-$0.48 &      & 32.5 & 30.6 &      &      & 31.9 &      \\
        &      &         &      &      &      &      &      &      &      \\
Ni I    &      &         &      &      &      &      &      &      &      \\
        &      &         &      &      &      &      &      &      &      \\
6108.12 & 1.68 & $-$2.49 & 39.9 & 43.3 & 39.5 & 32.2 & 33.2 & 51.1 &      \\
6767.78 & 1.83 & $-$2.11 & 47.1 & 61.0 & 61.3 & 51.5 & 51.3 & 52.0 & 49.3 \\
        &      &         &      &      &      &      &      &      &      \\
Ba II   &      &         &      &      &      &      &      &      &      \\
        &      &         &      &      &      &      &      &      &      \\
6141.75 & 0.70 &  ~~0.00 &112.3 &121.0 &118.5 &111.3 &112.4 &129.1 &135.4 \\
6496.91 & 0.60 & $-$0.38 &105.5 &124.9 &116.2 &117.1 &126.9 &122.2 &118.9 \\
\hline
\end{tabular}
\label{t:ewidths}
\end{table*}

\subsection{Atmospheric Parameters}

Effective temperatures were derived from dereddened $B-V$\ colour using the
calibration by Alonso et al. (1999): we interpolated the values at
[Fe/H]=$-1.5$\ from the tables for [Fe/H]=$-1$\ and $-2$. Rather than using
directly the individual stellar colours, we preferred to use the colours of
the mean loci at the same magnitudes of the program stars. Individual stellar
colours would have produced larger individual errors (0.014 mag,
correspondingto $\pm 34$~K, rather than $<\pm 10$~K with the procedure adopted
here: see below).

Internal uncertainties in these temperatures can be obtained by considering
the errors in the $V$\ magnitudes ($<0.05$~mag) and the slope of the magnitude
temperature relationship, which is 193~K/mag in the range of interest for the
program stars. We get internal uncertainties of $<10$~K, corresponding to
about 0.011~dex in [Fe/H]. Systematic errors are larger. The uncertainty in
the reddening ($\pm 0.005$~mag) multiplied for the slope of the
colour-temperature relation (about $-2400$~K/mag) yields a systematic error of
$\pm 12$~K, that is about 0.013 dex in [Fe/H]. Much larger is the uncertainty
in the adopted temperature scale, that is likely in the range 50-100~K,
producing possible errors in the Fe abundances in the range 0.055-0.11 dex.

Surface gravities were obtained from the location of the stars in the
colour-magnitude diagram. This procedure requires assumptions about the
distance modulus (from Gratton et al. 2003a), the bolometric corrections (from
Alonso et al. 1999), and the masses (we assumed a mass of 0.9~M$_\odot$, close
to the value given by isochrones fittings). Uncertainties in these gravities
are small (we estimate a total error of about 0.15 dex, dominated by
systematic effects in the temperature scale).

Microturbulent velocities $v_t$\ were determined by eliminating trends in the
relation between expected line strength and abundances (see Magain 1984). To
estimate errors in these values we notice that we found that the error in the
EWs contributes for 57\% of the variance of the errors in the abundances for
individual lines. Given the typical uncertainties in the slope of expected
line strength vs abundances, this implies an expected random error in the
microturbulent velocities of $\pm 0.09$~km s$^{-1}$. This value coincides with
the star-to-star scatter in microturbulent velocities.

Finally, model metal abundances were set in agreement with derived Fe
abundance. The adopted model atmosphere parameters are listed in
Table~\ref{t:uvesatmo}.

\begin{table}
\begin{center}
\caption{Atmospheric Parameters for stars observed with UVES}
\begin{tabular}{lcccc}
\hline
Star &T$_{\rm eff}$ &$\log g$ &$[$A/H$]$ &$v_t$\\
\hline
25072 & 5033 & 2.48 & $-1.49$ & 1.50 \\
26529 & 4888 & 2.12 & $-1.49$ & 1.50 \\
30409 & 4911 & 2.17 & $-1.49$ & 1.40 \\
30426 & 5009 & 2.42 & $-1.49$ & 1.30 \\
34854 & 4937 & 2.24 & $-1.49$ & 1.45 \\
37999 & 4854 & 2.02 & $-1.49$ & 1.50 \\
39672 & 4929 & 2.22 & $-1.49$ & 1.45 \\
\hline
\end{tabular}
\label{t:uvesatmo}
\end{center}
\end{table}

\subsection{Fe Abundances}

Individual [Fe/H] values are listed in Table ~\ref{t:uvesfe}, as well as
averages over the whole sample. Reference solar abundances are as in Gratton
et al. (2003b).

\begin{table}
\begin{center}
\caption{Iron abundances for stars observed with UVES}
\begin{tabular}{lcccccc}
\hline
Star  &    &   Fe I  &      &   &  Fe II  &     \\
      & n  &  [Fe/H] & r.m.s.  & n & [Fe/H]  & r.m.s. \\
\hline
25072 & 26 & $-1.46$ & 0.10 & 1 & $-1.59$ &      \\     
26529 & 26 & $-1.50$ & 0.08 & 3 & $-1.53$ & 0.05 \\
30409 & 25 & $-1.51$ & 0.10 & 2 & $-1.59$ & 0.09 \\
30426 & 25 & $-1.46$ & 0.11 & 2 & $-1.51$ & 0.04 \\ 
34854 & 23 & $-1.54$ & 0.15 & 1 & $-1.56$ &      \\
37999 & 31 & $-1.45$ & 0.08 & 3 & $-1.53$ & 0.05 \\
39672 & 16 & $-1.43$ & 0.24 &   &         &      \\
\hline
\multicolumn{7}{l}{$<{\rm [Fe/H] I}> = -1.48 \pm 0.01$}\\
\multicolumn{7}{l}{$<{\rm [Fe/H]II}> = -1.55 \pm 0.02$}\\
\hline
\end{tabular}
\label{t:uvesfe}
\end{center}
\end{table}

The average Fe abundance from all stars is  [Fe/H]=$-1.48\pm 0.02$\ (error of
the mean), with an r.m.s. scatter of 0.038 dex from 7 stars. If we consider
only the five stars with $S/N>25$, we have: [Fe/H]=$-1.48\pm 0.01$, with an
r.m.s. scatter of 0.028 dex.

There is a small offset of 0.07~dex between abundances given by neutral and
singly ionized Fe I lines. This might be attributed to the use of very few
lines for Fe II, but might also indicate some errors ($\sim 0.15$~dex) in the
surface gravities, as well as a systematic error of about 50~K in the
effective temperatures.

For comparison, Gratton et al. (2001) found [Fe/H]=$-1.44$ for neutral iron,
and $-1.55$ for singly ionized iron.

Table~\ref{t:errorparam} lists the impact of various uncertainties on the
derived Fe abundances. Variations in parameters of the model atmospheres 
(effective temperatures T$_{\rm eff}$, surface gravities $\log g$, model metal
abundances [A/H], microturbulent velocities $v_t$) were obtained by changing
each of the parameters at a time. The second column gives the variation of the
parameter used to estimate the changes in the abundances from neutral (Column
3) and singly ionized (Column 4) Fe lines. Columns 5 and 6 give the random
(i.e. appropriate to each star) and systematic (scale errors for all stars)
uncertainties in the various parameters; Columns 7 and 8 the corresponding
errors in the Fe abundances. The last row gives total errors: these have been
obtained by combining errors due to the various parameters.

\begin{table*}
\begin{center}
\caption{Uncertainties in Fe abundances for stars observed with UVES}
\begin{tabular}{lccccccc}
\hline
Parameter   & Variation  &$[$Fe/H$]$I&$[$Fe/H$]$II& Random   & Systematic & Total & Total \\
            &            &         &          & Error    &  Error   &Random & Syst. \\
\hline
EWs         &            &         &          &          &          & 0.013 & 0.006 \\ 
$\log gf$   &            &         &          &          &          &       & 0.019 \\
T$_{\rm eff}$ & 100 K    & ~~0.109 & $-$0.025 &   10 K   &   50 K   & 0.011 & 0.054 \\
$\log g$    & +0.3 dex   &$-$0.013 &  ~~0.125 & 0.02 dex & 0.15 dex & 0.000 & 0.006 \\
$[$A/H$]$   & +0.2 dex   & ~~0.003 & $-$0.013 & 0.03 dex & 0.06 dex & 0.001 & 0.001 \\
$v_t$       & +0.2 km s$^{-1}$  &$-$0.045 & $-$0.025 & 0.09 km s$^{-1}$& 0.06 km
s$^{-1}$& 0.020 & 0.013 \\
\\                         
Total       &            &         &          &          &          & 0.026 & 0.056 \\
\hline
\end{tabular}
\label{t:errorparam}
\end{center}
\end{table*}

\subsection{Intrinsic star-to-star scatter in Fe abundances}

The observed star-to-star scatter in Fe abundances is very small, in
particular if only higher quality ($S/N>25$) spectra are considered. In spite
of this, one may wonder if there is some evidence  for real star-to-star
scatter in the Fe abundances, or at least put some upper limit to this scatter
(even though the sample of stars observed with UVES is not extensive). In
Section 5.3 we have seen that the expected star-to-star scatter in Fe
abundances due to the adopted temperatures is only 0.008 dex. More relevant is
the error due to the microturbulent velocities.

To evaluate this source of error, we first note that considering only spectra
with $S/N>25$, the error in abundances from individual lines from each
spectrum is 0.068 dex, while the line-to-line r.m.s. scatter of the average
abundances from the 5 spectra is 0.097 dex. This indicates that only part of
the variance in the internal abundance errors is due to random errors in the
EWs, variable from star-to-star. The remaining contribution can be attributed
to systematic errors proper of each line (oscillator strengths, blends and
systematic effects on positioning of the continuum level). Given these facts,
the typical internal error in the abundances due to EWs can be estimated to be
$0.068/\sqrt{26}=0.013$~dex. Also, in the same way we may distribute the
measured errors in the microturbulent velocities (determined from the
$1-\sigma$\ uncertainty in the slope of the expected line strength vs
abundance fit) between random (i.e. star-to-star variable) errors, and
systematic (i.e. constant throughout the analysis of all stars) errors. Only
the random (star-to-star variable) error should be considered when discussing
the star-to-star abundance variations. By combining quadratically the various
sources of random errors, we get a prediction of 0.026 dex for the
star-to-star spread in the Fe abundances. This compares very well with the
measured star-to-star scatter of 0.028 dex. The conclusion is that there is
very scarce evidence for an intrinsic star-to-star scatter in the abundances;
a one-sided 1-$\sigma$\ upper limit is 0.017 dex.

\begin{figure}
\includegraphics[width=8.8cm]{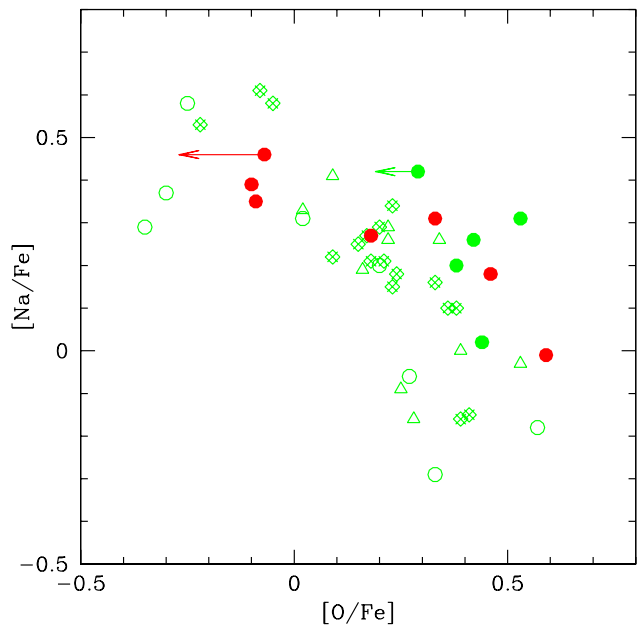}
\caption[]{[Na/Fe] ratio as a function of [O/Fe], for stars in NGC6752. Red
filled circles are our RGB bump stars in the present study. Green filled and
open circles are subgiant and turn-off stars, respectively, from Carretta et
al. (2005). Literature data are as follow: green diamonds with crosses inside
are bright red giants from the extensive study by Yong et al. (2003), open
green triangles are red giant stars from Norris and Da Costa (1995; 6 stars)
and Carretta (1994; 4 stars).}
\label{f:ona2}
\end{figure}

\subsection{O-Na anticorrelation}

The O-Na anticorrelation in NGC6752 is well known from previous observations
of both TO and subgiant stars (Gratton et al. 2001, Carretta et al. 2005), as
well as red giants (Carretta 1994; Norris \& Da Costa 1995; Yong et al. 2003).
It is fully confirmed by the present data for RGB bump stars (see
Figure~\ref{f:ona2}). Figure~\ref{f:ona2} collects also all data available up
to now, which clearly shows how an extensive  O-Na anticorrelation can be seen
along all evolutionary phases.

\subsection{Abundance of other elements}

Table~\ref{t:uvesother} lists the average abundances obtained for various
elements. Also in this case, solar abundances were as in Gratton et al.
(2003a). In the same Table we also compare the abundances obtained in this
paper with the analysis of Gratton et al. (2003a) and James et al. (2004). The
comparison is very good for the best determined elements: we found clear
overabundances of the $\alpha-$elements ([$\alpha$/Fe]=$+0.27\pm 0.01$), a
small deficiency of Ni, and slight overabundance of Ba. We notice that NGC6752
closely trace the composition of the dissipative component considered by
Gratton et al. (2003b), in agreement with its kinematics (Dinescu et al.
1999). The overabundance of $\alpha$-elements looks quite similar to those of
other globular clusters (see e.g. Gratton et al. 2004).

\begin{table*}
\begin{center}
\caption{Element-to-element abundance ratios for individual stars,
and averaged over the whole sample. For comparison, we also give the
values found by Gratton et al. (2003a) or James et al. (2004).}
\begin{tabular}{lcccccccccc}
\hline
Star  &n&[O/Fe]I&n&[Na/Fe]I&n&[Mg/Fe]I&n&[Si/Fe]I&n&[Ca/Fe]I\\
\hline
25072 & 1&$-$0.09&2&  +0.35& 2& +0.41 & 3& +0.34 & 9& +0.29 \\
26529 & 1&$-$0.10&2&  +0.39& 2& +0.41 & 2& +0.18 & 9& +0.37 \\
30409 & 1&  +0.46&2&  +0.18& 2& +0.60 & 3& +0.37 &10& +0.43 \\
30426 & 1&  +0.18&2&  +0.27& 2& +0.41 & 3& +0.22 &10& +0.38 \\
34854 & 1&  +0.33&2&  +0.31& 2& +0.69 & 3& +0.45 & 7& +0.43 \\
37999 & 1&  +0.59&1&$-$0.01& 2& +0.52 & 4& +0.32 &12& +0.34 \\
39672 &1&$<-$0.07&1&  +0.46& 2& +0.52 & 1& +0.31 & 6& +0.30 \\
\hline
Average& & +0.19&  & +0.29 &  & +0.51 &  & +0.31 &  & +0.36 \\
error  & &~~0.11&  &~~0.06 &  &~~0.04 &  &~~0.03 &  &~~0.02 \\
r.m.s. & &~~0.28&  &~~0.16 &  &~~0.11 &  &~~0.09 &  &~~0.06 \\
compare to &&    &  &       &  & +0.28 &  &       &  & +0.31 \\  
\hline
Star  &n&[Sc/Fe]II& n& [Ti/Fe]I&n&[Ni/Fe]I&n&[Ba/Fe]II&$[\alpha/Fe]$\\
\hline
25072 & 1&  +0.26 & 1& +0.30 & 2&$-$0.15& 2& +0.27 & +0.34 \\
26529 & 1&$-$0.04 & 1& +0.19 & 2&$-$0.14& 2& +0.29 & +0.29 \\
30409 & 1&$-$0.01 & 2& +0.15 & 2&$-$0.11& 2& +0.39 & +0.39 \\
30426 &  &        & 2& +0.17 & 2&$-$0.18& 2& +0.45 & +0.30 \\
34854 & 1&  +0.05 &  &       & 2&$-$0.21& 2& +0.37 & +0.44 \\
37999 & 2&$-$0.04 & 2& +0.16 & 2&$-$0.25& 2& +0.28 & +0.34 \\
39672 &  &        &  &       & 1&$-$0.32& 2& +0.43 & +0.33 \\
\hline
Average&  & +0.06  &  & +0.19 &  &$-$0.19&  & +0.35 & +0.35 \\
error  &  &~~0.05  &  &~~0.03 &  &~~0.03 &  &~~0.03 &~~0.02 \\
r.m.s. &  &~~0.12  &  &~~0.06 &  &~~0.07 &  &~~0.07 &~~0.05 \\
compare to &&$-$0.06& & +0.20 &  &$-$0.11&  & +0.18 & +0.27 \\  
\hline
\end{tabular}
\label{t:uvesother}
\end{center}
\end{table*}

\section{CONCLUSIONS}

We have shown that a single 1300~seconds exposure with FLAMES at VLT2 may
provide accurate estimates of the reddening toward NGC6752, as well as of the
chemical composition of the cluster (errors of 0.005~mag and of 0.02~dex
respectively). Similar analyses may provide results of comparable accuracy for
other globular clusters too on a uniform scale\footnote{Of course, scale
errors would be much larger. However, as mentioned in the Introduction, this
is not a serious concern if reddening and abundance scale are tied to those of
globular cluster by an appropriate consistent analysis as in Gratton et al.
(2003a)}. While results of similar accuracy have been already obtained for a
few clusters (including NGC6752), use of FLAMES allows to achieve such
accuracies with much less (about a factor of 20) observing time. An extensive
program over a large number of clusters may lead to large reductions of errors
in age determinations for those clusters for which accurate distances could be
obtained from either the main sequence fitting method, or even better from
dynamical methods.

\begin{acknowledgements}
This research has been funded by PRIN 2003029437 "Continuit\`a e
discontinuit\`a nella formazione della nostra Galassia" by Italian MIUR.
\end{acknowledgements}

\end{document}